\documentclass[12pt,preprint]{emulateapj}

\slugcomment{Submitted 2012 October 28, Accepted 2013 May 2}

\shorttitle{\textit{Suzaku} Study of  Bright Type I Seyfert Galaxy NGC 3516}
\shortauthors{Noda et al.}

\begin{document}


\title{\textit{Suzaku} Discovery of a Slowly Varying Hard X-ray Continuum \\ 
from the Type I Seyfert Galaxy NGC 3516}


\author{Hirofumi Noda\altaffilmark{1}, Kazuo Makishima\altaffilmark{2,3,4}, 
Kazuhiro Nakazawa\altaffilmark{2} and Shin'ya Yamada\altaffilmark{3}}

\altaffiltext{1}{Department of Astronomy, School of Science, The University of Tokyo}
\altaffiltext{2}{Department of Physics, School of Science, The University of Tokyo}
\altaffiltext{3}{Institute of Physical and Chemical Research (RIKEN)}
\altaffiltext{4}{Research Center for the Early Universe, University of Tokyo}


\begin{abstract}
The bright type I Seyfert galaxy NGC 3516 was observed 
by {\it Suzaku} twice, in 2005 October 12--15 and 2009 October 28--November 2, 
for a gross time coverage of 242 and 544 ksec and a net exposure of 134 and 255 ksec, respectively.
The 2--10 keV luminosity was $2.8 \times 10^{41}$ erg s$^{-1}$ in 2005, 
and $1.6 \times 10^{41}$ erg s$^{-1}$ in 2009. 
The 1.4--1.7 keV and 2--10 keV count rates both exhibited peak-to-peak variations
by a factor of $\sim2$ in 2005, while $\sim 4$ in 2009. In either observation, 
the 15--45 keV count rate was less variable. 
The 2--10 keV spectrum in 2005 was significantly more convex than that in 2009. 
Through a count-count-plot technique,
the 2--45 keV signals in both data were successfully decomposed 
in a model-independent way
into two distinct broadband components.
One is a variable emission with a featureless spectral shape,
and the other is a non-varying hard component accompanied by 
a prominent Fe-K emission line at  6.33 keV (6.40 keV in the rest frame).
The former was fitted successfully 
by an absorbed power-law model, 
while the latter requires a new hard continuum in addition to a reflection 
component from distant materials.  
The spectral and variability differences between the two observations are mainly 
attributed to long-term changes of this new hard continuum, which was stable on 
time scales of several hundreds ksec.

\end{abstract}





\section{Introduction}

In X-ray signals from Active Galactic Nuclei (AGNs), 
the primary continuum is presumably generated in a corona 
by the inverse Compton-scattering process (e.g., Haardt et al. 1994). 
A part of this emission region is sometimes covered by absorbers (either neutral or ionized), 
to produce so-called partial absorption condition (e.g, Holt et al. 1980; Miller et al. 2008).
When the primary emission is Compton-scattered or photo-absorbed in 
materials surrounding the central black hole (BH), 
a reflection component is generated (George \& Fabian 1991). 
Some of these secondary photons that are 
generated in central regions are subject to relativistic effects of the central BH, 
and appear as a relativistic reflection (Fabian \& Miniutti 2005). 
Thus, a typical X-ray spectrum of an AGN 
has been interpreted as a mixture of a primary continuum often modified 
by partial absorption, a distant neutral reflection with a narrow Fe-K line,
and a relativistically-blurred ionized reflection with a broad Fe-K line. 

In the above consensus view, the primary continuum from ``central engine'' has been assumed 
as a single Power Law (PL), while any spectral feature deviating from this 
modeling has been interpreted as due to its modification (e.g., by a partial 
absorber) or reprocessing (e.g.,  relativistically-blurred reflection).
This simplification, equivalent to an assumption of a single homogeneous corona, 
is necessitated by heavy degeneracy of various spectral components, 
particularly in hard X-ray bands which lack sharp spectral features.  
However, the central engine in reality would be significantly more complex, 
considering, e.g., strong radial gradients in the gravity and in physical conditions 
of the accreting matter. 
Then, timing information becomes important to identify 
individual spectral components and overcome this ambiguity, 
because variations are expected to differ from one component to another. 
Therefore, the AGN variability has long been employed in attempts to decompose
the overall emission into different components.

To extract the main variable component incorporating timing information, 
we can, e.g.,  employ the difference spectrum analysis method,
subtracting spectra between high-intensity and low-intensity periods
(e.g., Miniutti et al. 2007; Noda et al. 2011a). 
When the emission is considered to include
several variable components, 
the principle component analysis is useful 
(e.g., Miller et al. 2008; Noda et al. 2011a). 
These studies have revealed 
that the main variable component of many AGNs,
which is usually regarded as constituting their primary emission,
takes indeed a form of a single power-law (PL) 
of photon index $\sim2$ (e.g., Risaliti \& Elvis 2004). 
Although this apparently justifies the use of a single PL 
with a high-energy cutoff 
to approximate the primary continuum from the central engine, it is not obvious 
whether the primary emission as a 
whole can be represented by this variable component.

Variability-assisted studies of stable (or gradually varying) components, 
in principle, are often more difficult.
The popular root-mean-square variability analysis 
(e.g., Nandra et al. 1997; Markowitz et al. 2003)
does not allow us to distinguish 
whether an energy-dependent relative variation is caused, 
e.g., by the presence of constant signals in some energy bands,
or by shape changes in the variable component(s).
Similarly, the reverberation technique,
which can tell us the distance between the primary continuum source 
and the reflecting materials (e.g. Fabian et al. 2009; Zoghbi et al. 2012),
becomes ineffective if the primary variation is smeared out 
by light travel delays across the reflector 
so that the reflected signals lose time variability. 
As a result, the nature and composition of stable signals from AGNs 
have remained much less understood. 

In the present paper, we employ an intensity-assisted timing analysis 
called Count-Count Correlation with Positive Offset (C3PO) method, 
which was first developed to extract variable and stable signals from an X-ray spectrum of 
the leading black hole binary Cygnus X-1 (Churazov et al. 2001), 
and later tried on Seyferts (Taylor et al. 2003).
Applying this method to a soft X-ray band of \textit{Suzaku} data,
Noda et al. (2011b) and Noda et al. (2013) successfully revealed 
that the soft X-ray excess phenomena, 
widely seen in various types of disk-dominated AGNs,
originate at least in some cases as a relatively stable emission 
produced via thermal Comptonization  
in a corona that differs from the PL-generating one.
When applied to a harder/broader X-ray band, 
this method is expected to allow us to decompose broad-band spectra of AGNs 
into variable and stationary parts; hare, the latter will include the cold reflection 
component generated at large distances  
from the central BH, and possibly an additional new primary component 
as revealed by Noda et al. (2011a) in the hard (3--45 keV) band of MCG--6-30-15. 
Because the C3PO method is suited to AGNs with large X-ray variation amplitudes, 
we chose, in the present paper, the typical and bright type I Seyfert galaxy 
NGC 3516 at a redshift of $z=0.00885$, and analyzed 
archival \textit{Suzaku}  data of this AGN acquired on two occasions.
Unless otherwise stated, 
errors in the present paper refer to 90\% confidence limits.

\section{Observation}

NGC 3516 was observed by \textit{Suzaku} twice; 
first on 2005 October 12--15 during the Performance Verification phase, 
and the other on 2009
October 28--November 2 based on an AO7 key project 
that focused on broad Fe-K$\alpha$ emission lines. 
The XIS and HXD onboard \textit{Suzaku} were operated in their normal 
modes on both these occasions, 
and the source was placed at the XIS and HXD nominal 
positions in 2005 and 2009, respectively. 
The 2005 observation had a gross time coverage of 242 ksec, 
with a net exposure of  134 ksec with the XIS and 123 ksec with the HXD;
those of 2009 were 544 ksec, 251 ksec, and 191 ksec, respectively.

The 2005 \textit{Suzaku} data were already utilized by Markowitz et al. (2008), 
who reported the presence of a broad iron line with complex absorption. 
Patrick et al. (2011) analyzed the 2009 data,  
and found  no strong requirements for extremely broadened Fe-K lines. 
Analyzing the two \textit{Suzaku} data sets, together with those with \textit{XMM-Newton}, 
Turner et al. (2011) reported negative hard lags (unlike many Seyferts)
and variable soft X-ray absorption. 
According to these reports, the 2--10 keV intensity of NGC 3516 
varied during the \textit{Suzaku} observations by a factor of 2 or more.  
This makes both these datasets appropriate for the C3PO method.

In the present paper, the XIS and HXD-PIN data prepared via 
version 2.0 and 2.4 processing were utilized for the 2005 and 2009 observations, respectively. 
We added the data from XIS 0, 2, and 3 of 2005, 
and XIS 0 and 3 of 2009, and refer to them as XIS FI data, 
while we did not use those from XIS 1. 
On-source XIS FI events were accumulated over a circular region 
of $120''$ radius centered on the source, 
while background events were taken from a surrounding annular region, 
with the inner and outer radii of $180''$ and $270''$, respectively. 
The response matrices and ancillary response files were made by \texttt{xisrmfgen} and 
\texttt{xissimarfgen} (Ishisaki et al. 2007), respectively. 
The HXD-PIN events were prepared in a similar way. 
Non X-ray Background (NXB) included in the HXD-PIN data 
was estimated by analyzing fake events created 
by a standard NXB model (Fukazawa et al. 2009), 
and contribution from the Cosmic X-ray Background (Boldt et al. 1987) 
was calculated based on 
the spectral CXB brightness model, 
$9.0 \times 10 ^{-9} (E/3~\mathrm{keV})^{-0.29} \exp (-E/40~\mathrm{keV})$
erg cm$^{-2}$ s$^{-1}$ str $^{-1}$ keV$^{-1}$ (Gruber et al. 1999). 
They were then subtracted from the on-source events. 
In the present paper, we do not use the HXD-GSO data in either observation.

\section{Traditional Timing Analysis}

\subsection{Light curves and root mean square spectra}

\begin{figure*}[t]
\epsscale{1}
\plotone{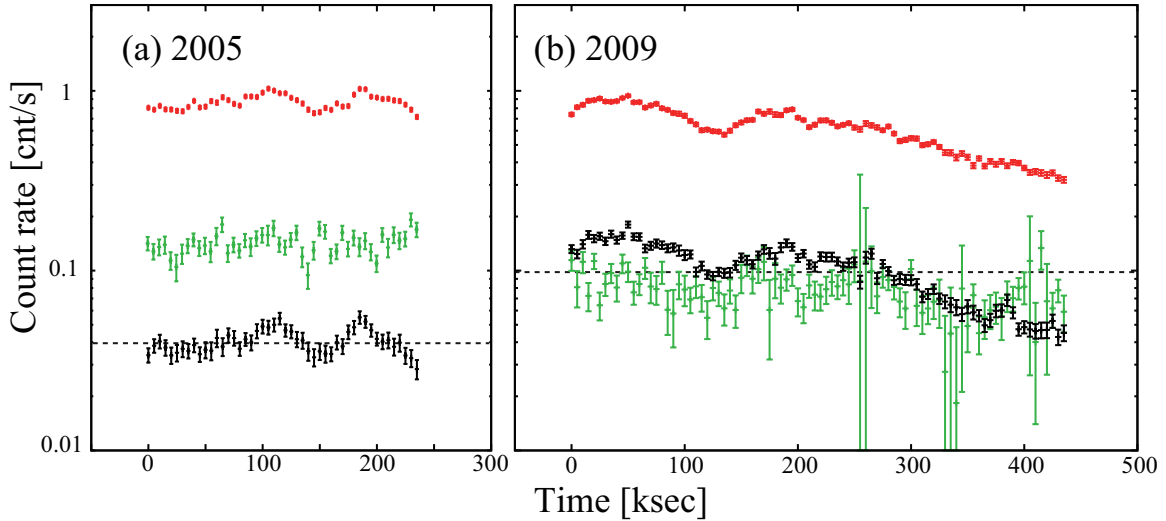}
\caption{Background-subtracted and dead-time corrected light curves of NGC 3516 
observed in 2005 (panel a) and 2009 (panel b). 
Both were measured with XIS FI in the 1.4--1.7 keV (black) and 1.7--10 keV (red) bands, 
together with the 15--45 keV HXD-PIN data (green), shown with a binning of 5 ksec. 
Error bars represent statistical $1\sigma$ ranges. The XIS count rate refers to
two cameras. 
A dashed line shows the average 1.4--1.7 keV count rate of $\sim 0.04$ cnt s$^{-1}$ in (a) and 
$\sim 0.1$ cnt s$^{-1}$ in (b). The time origin of the 2005 and 2009 observation are 
October 12 13:57:09 (UT) and October 28 5:35:15 (UT), respectively, 
and the end time are October 15 9:07:24 (UT) and 
November 2 12:39:14 (UT), respectively.  }
\end{figure*}

\begin{figure*}[t]
\epsscale{0.6}
\plotone{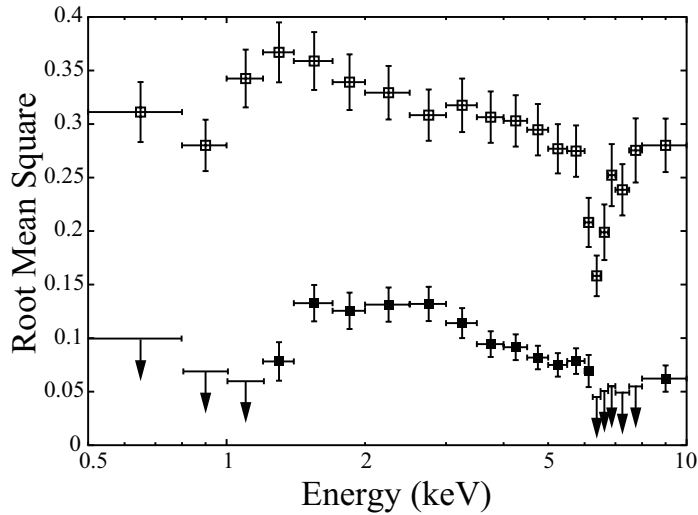}
\caption{The 0.5--10 keV RMS spectra from the 2005 (filled squares) and 2009 (open squares) 
observations.}
\end{figure*}

\begin{figure*}[t]
\epsscale{1}
\plotone{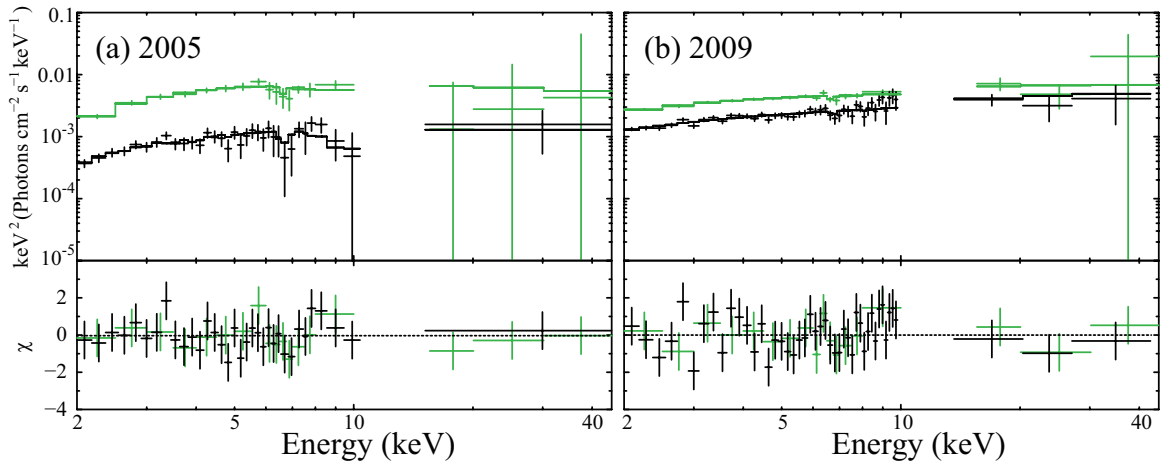}
\caption{The difference spectra (black) and the C3PO-derived variable components (green), 
in the 2005 (panel a) and 2009 (panel b) observation, 
presented in their deconvolved $\nu F_{\nu}$ form.
The fitted model is commonly \texttt{model\_v = wabs0 * zxipcf * cutoffPL0} (see text).}
\end{figure*}

\renewcommand{\arraystretch}{0.8}
\begin{table*}[t]
 \caption{Results of the fits to the difference spectra and the C3PO-derived variable spectra.}
 \label{all_tbl}
 \begin{center}
  \begin{tabular}{ccccccc}
   \hline\hline
   & &   \multicolumn{2}{c}{Difference} &  \multicolumn{2}{c}{Variable$^{\rm a}$}\\
   Component & Parameter & 2005 & 2009 & 2005 & 2009 \\
          \hline
                      	 
     \texttt{wabs0} &$ N_{\rm H0}^{\rm b}$
			& $3.0 \pm 1.4$
			& $0.6 \pm 0.5$
			& $3.8^{+3.8}_{-1.3}$
			& $0.9 \pm 0.4$\\

     \texttt{zxipcf} &$ N_{\rm i}^{\rm b}$
			&$89.1^{+90.4}_{-85.6}$
                          &$<254.2$
                          &$<57.4$
                          &$<2.1$ \\
                          
                          &$\log\xi$
			&$3.55^{+2.09}_{-1.42}$
                          &$ <4.40$
                          &$ <3.28$
                          &$<3.55$ \\
                         
  \texttt{cutoffPL0} & $\Gamma_0$
  			& $2.05^{+0.49}_{-0.43}$
			& $1.59 \pm 0.12$
			& $2.12^{+0.41}_{-0.28}$
			& $1.75^{+0.06}_{-0.05}$\\\
			
			& $E_{\rm cut}$~(keV)
			&  \multicolumn{4}{c}{150 (fix)}\\
			
			&$N_{\rm PL}^{\rm c}$
			&$1.67^{+6.11}_{-1.02}$
			&$1.25^{+0.31}_{-0.25}$
			&$0.92^{+1.43}_{-0.38}$
			&$0.31 \pm 0.06$\\[1.5ex]

   $\chi^{2}$/d.o.f. & & 17.5/29 & 40.9/41  & 8.0/13 & 9.5/13  \\
\hline\hline

  \end{tabular}
\end{center}
   	{\small
	$^{\rm a}$ The variable spectra refer to the case with $C=0$.\\
	$^{\rm b}$ Equivalent hydrogen column density in  $10^{22}$ cm$^{-2}$. \\
         $^{\rm c}$ The power-law normalization at 1 keV, in units of $10^{-3}$~photons~keV$^{-1}$~cm$^{-2}$~s$^{-1}$~at 1 keV.\\
         
}

\end{table*}


Figure 1 shows XIS FI (in 2 bands) and HXD-PIN 
light curves of  NGC 3516 in 2005 and 2009. 
Here and hereafter, we present the XIS light curves  
as a sum of two cameras, with the 2005 counts (sum of three cameras) 
multiplied by a factor of $\sim1/2$ (including nominal-position correcting factor). 
When compared to the source intensity in 2009, that in 2005 was higher 
by a factor $\sim 1.5$ in the middle band, and $\sim 2$ in the HXD band, 
but $\sim 2.5$ times lower in the lowest energies. 
Therefore, the overall spectrum in 2005 is considered to 
be significantly more convex than that in 2009. 
Characteristics of time variations are also somewhat different between them. 
The 2005 light curves fluctuate around the average count rates, 
while those in 2009 show a more monotonic decrease. 
The 1.4--1.7 keV band count rate varied by 50\% (peak-to-peak) in 2005 
and a factor of $\sim 4$ in 2009.   
That of the 1.7--10 keV band is slightly smaller in both observations, 
and the 15--45 keV variation is even smaller. 

To investigate more quantitatively the soft X-ray variablity, 
a popular timing analysis method, Root Mean Square (RMS) analysis
(e.g., Nandra et al. 1997; Markowitz et al. 2003; Matsumoto \& Inoue 2003) 
was applied to the 0.5--10 keV XIS FI data. 
Specifically, we divided the 0.5--10 keV broad band into 21 finer bands 
with boundaries at 0.5, 0.8, 1.0, 1.2, 1.4, 1.7, 2.0, 2.5, 3.0, 3.5, 
4.0, 4.5, 5.0, 5.5, 6.0, 6.25, 6.5, 6.75, 7.0, 7.5, 8.0, and 10.0 keV. 
The energy intervals were thus set narrower across 
the Fe-K$\alpha$ line energy region (6.0--7.0 keV). 
The RMS spectra derived from the two observations are shown in Figure 2. 
As expected from the light curves, the 2005 variability was 
less than $\sim 30$\% of that in 2009. 
However, the RMS spectral shape is similar between the two in energies above 2.5 keV.  
The most variable band in both observations is around 1.4--1.7 keV, 
while the least variable one is $\sim 6.33$ keV due to 
an Fe-K$\alpha$ line (at 6.40 keV in the rest frame), 
which is thus inferred to be less variable than the continuum. 
Below 1.4 keV, the two RMS spectra both decrease presumably because of  
the presence of stable signals including thin-thermal plasma emission 
from the host galaxy (George et al. 2002). 
To exclude such soft X-ray contaminants, 
we hereafter choose the most variable 1.4--1.7 band as a reference, 
and study the source behavior in energies above 2 keV.

\subsection{Difference spectrum analysis}

To conventionally extract variable components,  
we applied the difference spectrum analysis (\S1) to the two \textit{Suzaku} data sets.
As shown in Figure 1 by a dotted line, 
we divided the whole observation into two phases in which the 1.4--1.7 keV count rate 
is higher (denoted High-phase) and lower (Low phase) than the average. 
Then, the spectrum accumulated over the Low phase was subtracted from 
that over the High phase, to obtain a difference spectrum. 

Figure 3 (black) shows the difference spectra thus derived from the 2005 and 2009 data. 
First, we fitted them with an absorbed cutoff PL model, \texttt{wabs0 * cuotffPL0}, 
where \texttt{wabs0} represents a sum of  the intrinsic and the Galactic absorption, 
to be applied hereafter to all model components from the AGN. 
The fits were both acceptable with $\chi^2$/d.o.f.$=21.4/31$ in 2005 
and $\chi^2$/d.o.f.$=42.2/43$ in 2009, 
and gave the column density $N_{\rm H0}$ of \texttt{wabs0} and 
the photon index $\Gamma_0$ of \texttt{cutoffPL0} as 
$2.7^{+1.4}_{-1.3} \times 10^{22}$ cm$^{-2}$ and $1.91^{+0.42}_{-0.36}$ in 2005, respectively, 
while $0.7^{+0.5}_{-0.2} \times 10^{22}$ cm$^{-2}$ and $1.60^{+0.13}_{-0.12}$ in 2009, respectively. 
The neutral column density is thus significantly higher in 2005 than in 2009, 
while the spectral shape is not significantly different between the two spectra.

Although the fits with the absorbed PL model are successful, 
we find negative residuals at $\sim 6.5-7$ keV,  
which can be identified with the ionized Fe absorption features reported by Turner et al. (2008, 2011). 
Thus, we refitted the difference spectra with a model of the form \texttt{model\_v}$\equiv$\texttt{wabs0 * zxipcf * cutoffPL0}, 
where \texttt{zxipcf} represents the ionized absorption. 
We left free the column density $N_{\rm H0}$ of \texttt{wabs0} and 
the photon index $\Gamma_0$ of 
\texttt{cutoffPL0}, and fixed the high energy cutoff at 150 keV. 
The column density $N_{\rm i}$ and the ionization parameter $\xi$ of \texttt{zxipcf} were left free, 
whereas its covering fraction was fixed at 1.0 for simplicity, and the redshift at 0.00885. 
The inclusion of the \texttt{zxipcf} factor, with two free parameters, 
improved the fits by $\Delta \chi^2 \sim 3.9$ for 2005 and $\sim 1.3$ for 2009, 
and yielded the parameters as given in Table 1. 
Although the significance of this additional feature is not compelling (with a chance occurrence probability 
of $\sim 16$\% in 2005 and $\sim54$ \% in 2009),
we hereafter retain this factor in view of consistency with the previous works. 

\begin{figure*}[p]
\epsscale{1}
\plotone{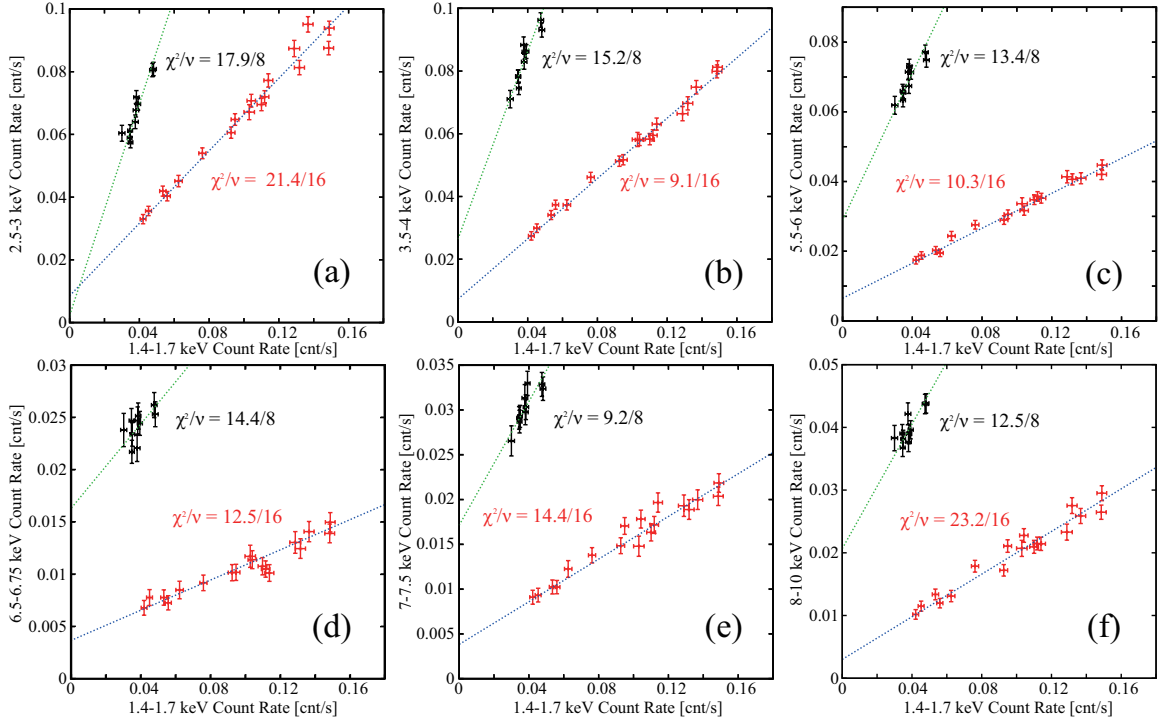}
\caption{Six CCPs of the 2005 (black) and 2009 (red) observations,  
in which abscissa gives NXB-subtracted XIS FI count rate (per two cameras) in 1.4--1.7 keV, 
while ordinate that in (a) 2.5--3 keV, (b) 3.5--4 keV, (c) 5.5--6 keV, (d) 6.5--6.75 keV, 
(e) 7--7.5 keV, and (f) 8--10 keV band count rates. All data are binned into 25 ksec
The 2005 data are shown after corrected for differences of the number of CCD cameras, 
the difference in pointing positions, and slight response changes between the two epochs.  
The error bars represent statistical $\pm 1 \sigma$ range. 
The dotted straight line refers to eq. (1). 
The $\chi^2$/d.o.f. values are shown in each panel in black (for 2005) and red (2009). }
\end{figure*}

\renewcommand{\arraystretch}{1}
\begin{table*}[h]
 \caption{Parameters (with 1$\sigma$ errors) obtained by fitting 16 CCPs with eq. (1). }
 \label{all_tbl}
 \small
 \begin{center}
  \begin{tabular}{cccccccc}
   \hline\hline

  &\multicolumn{3}{c}{2005} &&  \multicolumn{3}{c}{2009}  \\\hline
  Range (keV)  & $A$ & $B \times10$ & $B'_{\rm max} \times 10$& &$A \times 10$ & $B \times 10^{2}$ & $B'_{\rm max} \times 10^{2}$\\

   \hline
  2--2.5 & $1.42\pm0.10$ & $0.03\pm0.06$  & $0.42\pm0.02$& &$0.76\pm0.02$ & $0.06\pm0.02$  & $0.31\pm0.01$  \\[-0ex]
2.5--3 & $1.69\pm0.14$ & $0.02\pm0.08$  & $0.48\pm0.02$& &$0.58\pm0.02$ & $0.09\pm0.02$  & $0.27\pm0.01$  \\[-0ex]    
 3--3.5 & $1.69\pm0.15$ & $0.09\pm0.12$  & $0.59\pm0.03$& &$5.41\pm0.15$ & $0.62\pm0.14$  & $2.36\pm0.10$  \\[-0ex]
3.5--4   & $1.49\pm0.14$& $0.27\pm0.08$ &$0.67\pm0.02$ &&$4.81\pm0.18$& $0.73\pm0.16$ &$2.28\pm0.11$ \\[-0ex]
4--4.5 &  $1.41\pm0.11$& $0.30\pm0.06$  &$0.68\pm0.02$  &  &$4.21\pm0.16$& $0.67\pm0.15$  &$2.02\pm0.10$\\ [-0ex] 
4.5--5   & $1.26\pm0.14$ & $0.33\pm0.08$ &$0.67\pm0.03$& &$3.54\pm0.12$ & $0.66\pm0.13$ &$1.80\pm0.09$ \\[-0ex]
5--5.5   & $1.10\pm0.18$ & $0.35\pm0.10$ &$0.65\pm0.03$  &  &$2.98\pm0.13$ & $0.76\pm0.12$ &$1.72\pm0.08$ \\[-0ex]
5.5--6   & $1.05\pm0.12$ & $0.29\pm0.06$  &$0.57\pm0.02$& &$2.51\pm0.12$ & $0.64\pm0.12$  &$1.45\pm0.09$ \\[-0ex]
6--6.25    &$0.34\pm0.09$ & $0.22\pm0.05$ &$0.31\pm0.01$  & &$1.02\pm0.08$ & $0.51\pm0.08$ &$0.83\pm0.05$\\[-0ex]
6.25--6.5    &  $0.26\pm0.09$ & $0.32\pm0.05$ & $0.39\pm0.01$& & $1.11\pm0.11$ & $1.10\pm0.11$ & $1.46\pm0.07$ \\ [-0ex]
6.5--6.75    & $0.20\pm0.07$ & $0.16\pm0.04$ &$0.22\pm0.01$ && $0.72\pm0.06$ & $0.35\pm0.05$ &$0.58\pm0.04$\\ [-0ex]
 6.75--7    &  $0.15\pm0.05$ & $0.15\pm0.03$ &$0.19\pm0.01$ && $0.57\pm0.08$ & $0.30\pm0.07$ &$0.48\pm0.05$\\ [-0ex]
7--7.5    & $0.35\pm0.04$ & $0.17\pm0.03$ &$0.26\pm0.01$  &  & $1.09\pm0.08$ & $0.48\pm0.08$ &$0.83\pm0.06$\\ [-0ex]
7.5--8    & $0.24\pm0.06$ & $0.11\pm0.04$ &$0.18\pm0.01$& & $0.79\pm0.06$ & $0.15\pm0.06$ &$0.40\pm0.38$ \\ [-0ex]
8--10    &  $0.49\pm0.08$ & $0.20\pm0.05$ &$0.34\pm0.01$ & & $1.56\pm0.10$ & $0.44\pm0.09$ &$0.94\pm0.06$ \\ [-0ex]
15--20    &  $0.06\pm0.20$ & $0.68\pm0.11$ &  $0.69\pm0.04$& & $1.41\pm0.32$ & $2.57\pm0.32$  & $3.03\pm0.23$ \\ [-0ex]
 20--30    & $0.10 \pm 0.30$ & $0.53\pm0.17$ & $0.55\pm0.06$  & &$0.74\pm0.31$ & $2.56\pm0.41$  &$2.78\pm0.32$  \\ [-0ex]
30--45    &$0.04\pm0.27$ & $0.16\pm0.15$ &$0.18 \pm0.05 $  & &$0.78\pm0.97$& $0.03\pm0.95$  & $0.25\pm0.68$\\                      
      \hline\hline

  \end{tabular}
 \end{center}

\end{table*}
\renewcommand{\arraystretch}{1.0}

\begin{figure*}[t]
\epsscale{1}
\plotone{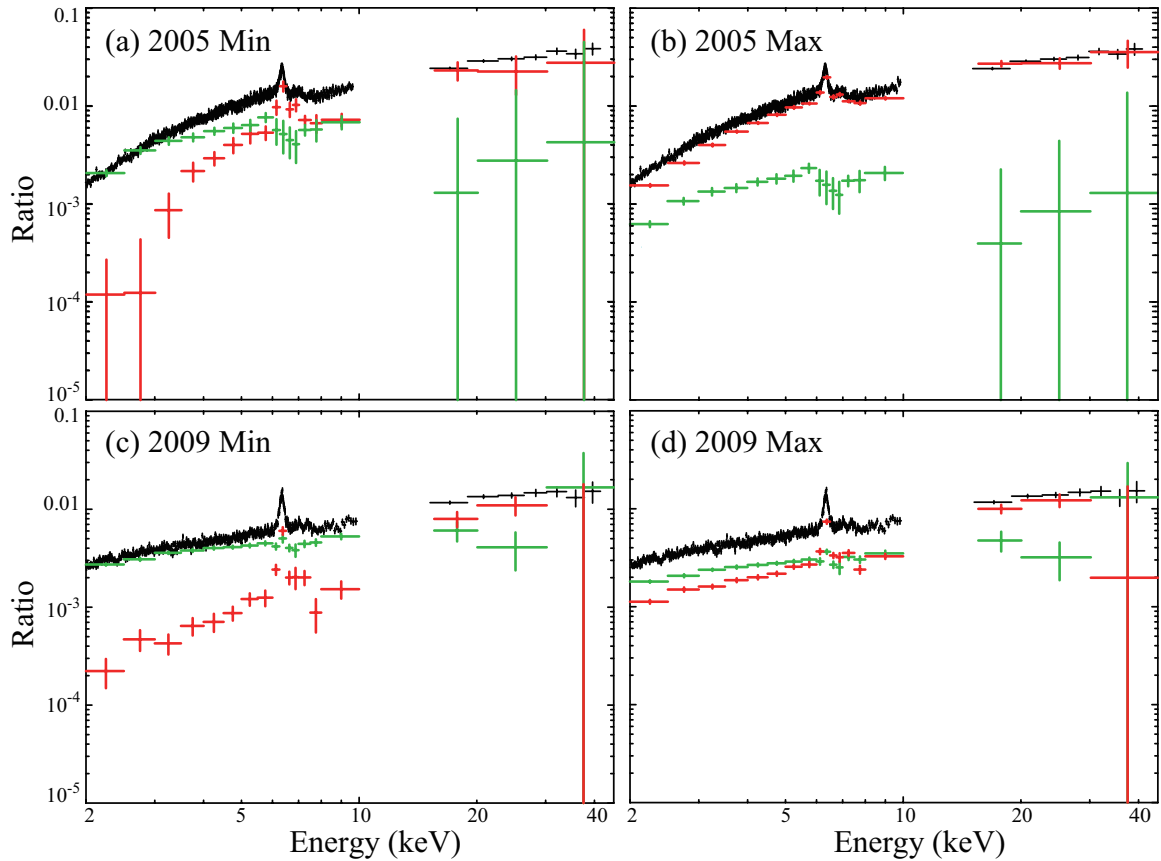}
\caption{Background-subtracted time-averaged spectra (black), 
and the C3PO-extracted variable (green) and stable (red) spectra of 2005 (panel a and b) 
and  2009 (c and d), 
all shown as ratios to a common PL of photon index 2. 
Panels (a) and (c) show the case of $C=0$ in eq. (2), while (b) and (d) those of $C=C_{\rm max}$. }
\end{figure*}

\begin{figure*}[p]
\epsscale{0.9}
\plotone{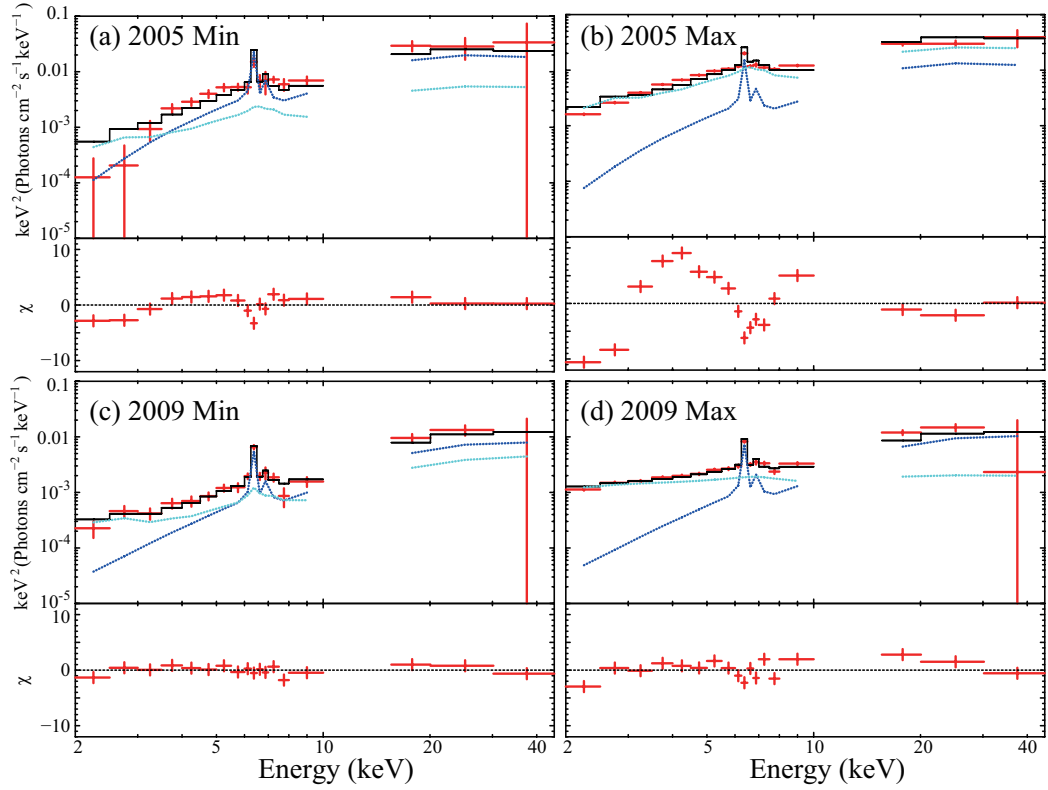}
\caption{The C3PO-derived stable component (red crosses), in $\nu F_{\nu}$ form, 
in (a) the 2005 min, (b) the 2005 max, (c) the 2009 min, and  (d) the 2009 max cases. 
They are fitted with the sum of a distant cold reflection (blue) and 
an ionized reflection (cyan), namely, \texttt{wabs0 * (pexmon + reflionx)}.  }
\end{figure*}

\begin{figure*}[t]
\epsscale{0.9}
\plotone{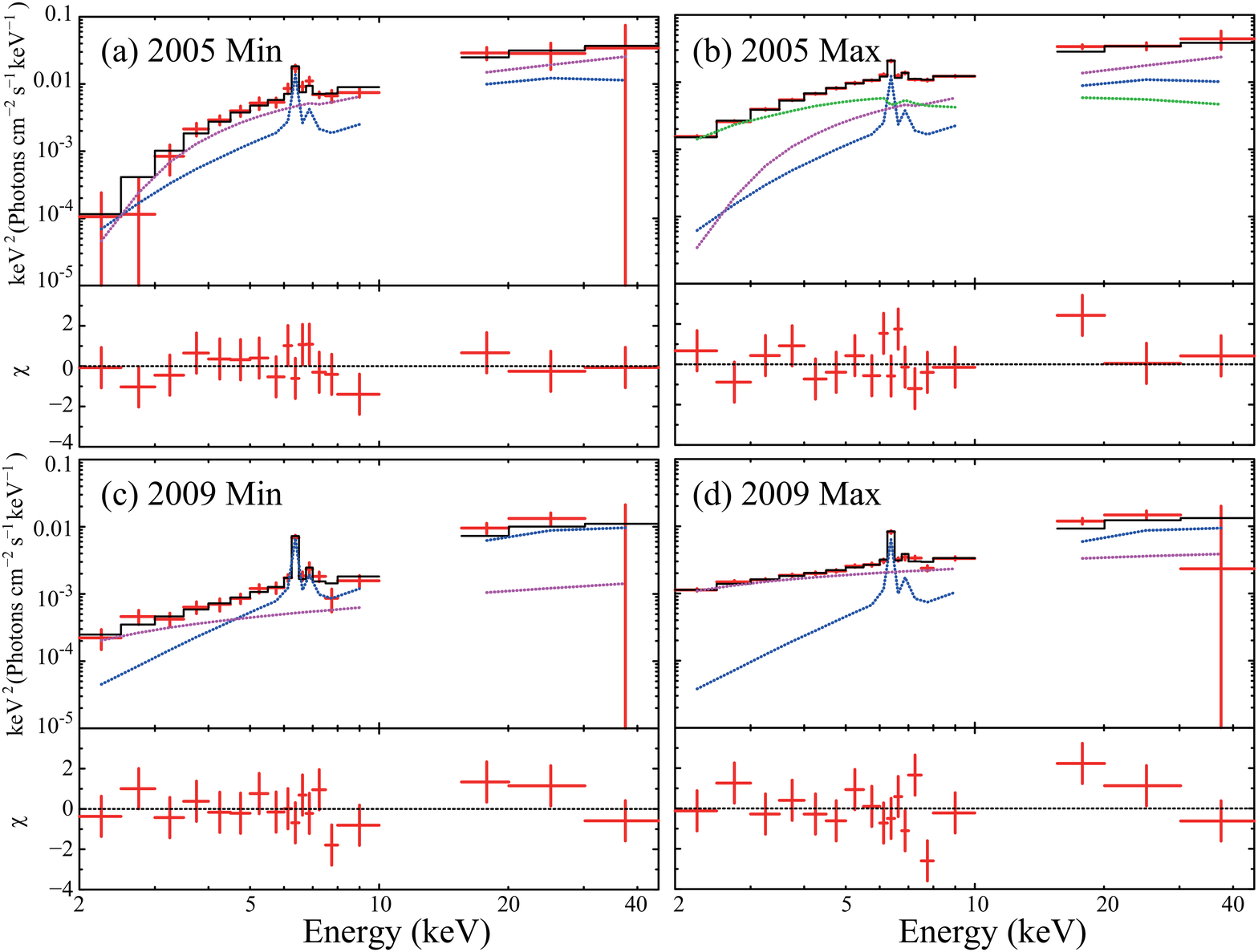}
\caption{Same as Figure 6, but the ionized reflection replaced with an absorbed cutoff PL model (purple). 
Namely,  the fitted model is \texttt{wabs0 * (pexmon + wabs1 * cutoffPL1)}, 
except in (b) where \texttt{wabs * zxipcf * cutoffPL2} is added. }
\end{figure*}

\section{Novel Timing Analysis}

\subsection{The count-count correlation with positive offset}

To more systematically decompose the 3--45 keV emission into
the variable and stationary  components,  
the C3PO method (Noda et al. 2011b; Noda et al. 2012) provides a powerful tool. 
We chose the 1.4--1.7 keV band as a reference, 
because the variability was largest therein in both data sets. 
We divided the 3--10 keV XIS band into 13 finer bands 
with the same boundaries as the RMS spectra, 
while the 15--45 keV HXD-PIN data into 3 bands with boundaries at 
15.0, 20.0, 30.0, and 45.0 keV. 
Then, as shown in Figure 4, we made  16 Count-Count Plots (CCPs), 
in which ordinate (denoted $y$) gives NXB-subtracted 
XIS FI or HXD-PIN count rates in these bands, while abscissa (denoted $x$) 
is those in the 1.4--1.7 keV band used as the reference. 
The CCPs all exhibit a linear correlation, 
but those in 2009 have much larger variation amplitudes than those of 2005, 
as expected from the light curves (Figure 1) and the RMS spectra (Figure 2).
Compared to the 2009 CCPs, 
those in 2005 show much steeper slopes, and larger y-intercepts.

\renewcommand{\arraystretch}{0.3}
\begin{table*}[t]
 \caption{Results of spectral fits to the stable spectra derived by the C3PO method$^{\rm a}$.}
 \label{all_tbl}
 \begin{center}
  \begin{tabular}{ccccccc}
   \hline\hline
   &  &   \multicolumn{2}{c}{2005} & \multicolumn{2}{c}{2009}\\
   Component &Parameter & min & max & min & max\\
   
   \hline
		
    \texttt{wabs0} &$ N_{\rm H0}^{\rm b}$
			&   \multicolumn{2}{c}{3.6 (fix)} 
                          &  \multicolumn{2}{c}{0.9 (fix)}  \\[1.2ex]

    \texttt{pexmon} & $\Gamma_\mathrm{ref}$%
    			& \multicolumn{2}{c}{2.12 (fix)}
                          & \multicolumn{2}{c}{1.75 (fix)}\\
                          
                          & $E_{\rm cut}$~(keV)
                      & \multicolumn{4}{c}{150 (fix)}\\
                          
                          &$A_{\rm Fe}$~($Z_{\odot}$)
                          & \multicolumn{4}{c}{1 (fix)}\\
                          
                         &$f_{\rm ref}$~($\Omega/2\pi$)
                          & \multicolumn{4}{c}{1 (fix)}\\
                          
                         &$I$~(degree)
                          & \multicolumn{4}{c}{60 (fix)}\\
                          
                          & $N_\mathrm{ref}^{\rm c}$%
    			&$0.72$		
			&$13.42$
                          & $0.86$
                          &$1.48$\\[1.5ex]
   $\chi^2$/d.o.f. &  & 66.07/17& 4802.81/17  & 55.46/17& 1903.97/17\\

   \hline    
   
       \texttt{wabs0} &$ N_{\rm H}^{\rm b}$
			&   \multicolumn{2}{c}{3.6 (fix)} 
                          &  \multicolumn{2}{c}{0.9 (fix)}  \\[1.2ex]

    \texttt{reflionx} & $\Gamma_\mathrm{ref}$%
    			& \multicolumn{2}{c}{2.12 (fix)}
                          & \multicolumn{2}{c}{1.75 (fix)}\\
                                                    
                          &$A_{\rm Fe}$~($Z_{\odot}$)
                          & \multicolumn{4}{c}{1 (fix)}\\

                         & $\xi$ (erg cm s$^{-1}$)
			&$14.1$
			& $125.9$
                          &$58.32^{+8.6}_{-5.4}$
                          &$164.7$\\

                          & $N_\mathrm{ref}^{\rm d}$%
    			&$67.94$		
			&$8.63$
                          & $3.78 \pm 0.70$
                          &$1.93$\\[1.5ex]
   $\chi^2$/d.o.f. &  & 48.64/16& 775.23/16  & 23.30/16& 570.13/16\\

   \hline

     \texttt{wabs0} &$ N_{\rm H}^{\rm b}$
			&   \multicolumn{2}{c}{3.6 (fix)} 
                          &  \multicolumn{2}{c}{0.90 (fix)}  \\[1.2ex]

    \texttt{pexmon} & $\Gamma_\mathrm{ref}$%
    			& \multicolumn{2}{c}{2.12 (fix)}
                          & \multicolumn{2}{c}{1.75 (fix)}\\
                          
                          & $E_{\rm cut}$~(keV)
                      & \multicolumn{4}{c}{150 (fix)}\\
                          
                          &$A_{\rm Fe}$~($Z_{\odot}$)
                          & \multicolumn{4}{c}{1 (fix)}\\
                          
                         &$f_{\rm ref}$~($\Omega/2\pi$)
                          & \multicolumn{4}{c}{1 (fix)}\\
                          
                         &$I$~(degree)
                          & \multicolumn{4}{c}{60 (fix)}\\
                          
                          & $N_\mathrm{ref}^{\rm c}$%
    			&$3.85$
			&$3.33$
			&$0.56^{+0.10}_{-0.16}$
                          & $0.15$\\

 \texttt{reflionx} & $\Gamma_\mathrm{ref}$%
    			& \multicolumn{2}{c}{2.12 (fix)}
                           & \multicolumn{2}{c}{1.75 (fix)}\\

                          &$A_{\rm Fe}$~($Z_{\odot}$)
                          & \multicolumn{4}{c}{1 (fix)}\\

                         & $\xi$ (erg cm s$^{-1}$)
			&$199.5$
			& $199.9$
                          &$197.1^{+14.3}_{-72.4}$
                          &$199.9$\\

                          & $N_\mathrm{ref}^{\rm d}$%
			&$0.98$
    			& $4.26$
                            &$0.35^{+0.62}_{-0.16}$
                          & $1.49$\\[1.5ex]
                          
   $\chi^2$/d.o.f. & & 45.77/15& 532.23/15 &  9.52/15 & 78.99/15\\
        \hline\hline

  \end{tabular}
\end{center}
   	{\small
	$^{\rm a}$ The errors refer to 90\% confidence ranges. \\
	$^{\rm b}$ Equivalent hydrogen column density in  $10^{22}$ cm$^{-2}$. \\
         $^{\rm c}$ The \texttt{pexmon} normalization at 1 keV, in units of $10^{-2}$~photons~keV$^{-1}$~cm$^{-2}$~s$^{-1}$~at 1 keV.\\
          $^{\rm d}$ The \texttt{reflionx} normalization, in units of $10^{-6}$.}

\end{table*}
\renewcommand{\arraystretch}{1}

\renewcommand{\arraystretch}{0.3}
\begin{table*}[t]
 \caption{Same as Table 3, but with an empirical cutoff-PL or 
 a relativistically blurred refection included$^{\rm a}$.}
 \label{all_tbl}
 \begin{center}
  \begin{tabular}{ccccccc}
   \hline\hline
   &  &   \multicolumn{2}{c}{2005} & \multicolumn{2}{c}{2009}\\
   Component &Parameter & min & max & min & max\\
   
   \hline
      
     \texttt{wabs0} &$ N_{\rm H0}^{\rm b}$
			&   \multicolumn{2}{c}{3.6 (fix)} 
                          &  \multicolumn{2}{c}{0.90 (fix)}  \\[1.2ex]

    \texttt{pexmon} & $\Gamma_\mathrm{ref}$%
    			& \multicolumn{2}{c}{2.12 (fix)}
                          & \multicolumn{2}{c}{1.75 (fix)}\\
                          
                          & $E_{\rm cut}$~(keV)
                      & \multicolumn{4}{c}{150 (fix)}\\
                          
                          &$A_{\rm Fe}$~($Z_{\odot}$)
                          & \multicolumn{4}{c}{1 (fix)}\\
                          
                         &$f_{\rm ref}$~($\Omega/2\pi$)
                          & \multicolumn{4}{c}{1 (fix)}\\
                          
                         &$I$~(degree)
                          & \multicolumn{4}{c}{60 (fix)}\\
                          
                          & $N_\mathrm{ref}^{\rm c}$%
			&$1.18 \pm 0.50$    			
			&$1.25 \pm 0.31$
			&$0.49^{+0.09}_{-0.08}$
                          & $0.48^{+0.05}_{-0.06}$\\
   
    \texttt{wabs1} &$ N_{\rm H1}^{\rm b}$
			&$9.4^{+5.5}_{-5.2}$			
			&$10.2^{+3.3}_{-2.9}$
                          &$<4.31$			
			&$<0.49$\\
   
   \texttt{cutoffPL1}     & $\Gamma_1$%
   			&$1.18^{+0.74}_{-0.49}$
			&$1.11^{+0.24}_{-0.34}$
			&$1.46^{+1.48}_{-0.69}$
			&$1.67^{+0.17}_{-0.14}$\\
                      
                      & $E_{\rm cut}$~(keV)
                     & \multicolumn{4}{c}{150 (fix)}\\

                    & $N_\mathrm{PL}^{\rm d}$%
                     &$1.45^{+0.88}_{-0.37}$
                     &$1.03^{+0.43}_{-0.79}$
                       &$<0.04$                   
                     &$0.12^{+0.03}_{-0.02}$\\[1.2ex]
                     
        \texttt{zxipcf} &$ N_{\rm i}^{\rm b}$
			&--
			& $37.1^{+12.2}_{-11.6}$
			&--
                          &-- \\
                          
                          &$\log\xi$
			&--
			& $3.03^{+0.15}_{-0.07}$
			&--
                          &-- \\
                          
  \texttt{cutoffPL2} & $\Gamma_2$
  			&--
			& $2.22^{+0.26}_{-0.36}$
			&--
			&--\\
			
			& $E_{\rm cut}$~(keV)
			&--
			&150 (fix)
			&--
			&--\\
			
			&$N_{\rm PL}^{\rm d}$
			&--
			&$1.13^{+0.18}_{-0.13}$
			&--
			&--\\[1.2ex]

   $\chi^{2}$/d.o.f. & & 7.13/15 & 16.92/11 & 10.89/15  &21.40/15\\
      \hline
      
           \texttt{wabs0} &$ N_{\rm H0}^{\rm b}$
			&   \multicolumn{2}{c}{3.6 (fix)} 
                          &  \multicolumn{2}{c}{0.90 (fix)}  \\[1.2ex]

    \texttt{pexmon} & $\Gamma_\mathrm{ref}$%
    			& \multicolumn{2}{c}{2.12 (fix)}
                          & \multicolumn{2}{c}{1.75 (fix)}\\
                          
                          & $E_{\rm cut}$~(keV)
                      & \multicolumn{4}{c}{150 (fix)}\\
                          
                          &$A_{\rm Fe}$~($Z_{\odot}$)
                          & \multicolumn{4}{c}{1 (fix)}\\
                          
                         &$f_{\rm ref}$~($\Omega/2\pi$)
                          & \multicolumn{4}{c}{1 (fix)}\\
                          
                         &$I$~(degree)
                          & \multicolumn{4}{c}{60 (fix)}\\
                          
                          & $N_\mathrm{ref}^{\rm c}$%
    			&$2.23^{+1.19}_{-1.42}$
			&$2.54^{+0.39}_{-0.34}$
			&$<0.62$
                          & $0.57^{+0.10}_{-0.09}$\\[1.2ex]

 \texttt{kdblur} & $q$%
    			& $>3.2$
                           & $7.1^{+1.3}_{-1.1}$
                           &$>6.9$
                           &$>4.8$\\
			
			&$R_{\rm in}$
			&$<2.04$
			&$<1.7$
			&$<3.3$
			&$<1.5$\\

 \texttt{reflionx} & $\Gamma_\mathrm{ref}$%
    			& \multicolumn{2}{c}{2.12 (fix)}
                           & \multicolumn{2}{c}{1.75 (fix)}\\

                          &$A_{\rm Fe}$~($Z_{\odot}$)
                          & \multicolumn{4}{c}{1 (fix)}\\

                         & $\xi$ (erg cm s$^{-1}$)
			&$<16.3$
			& $<73.6$
                          &$59.4^{+429.1}_{-29.8}$
                          &$<361.5$\\

                          & $N_\mathrm{ref}^{\rm e}$%
			&$0.76^{+0.19}_{-0.36}$
    			& $59.93^{+21.31}_{-18.69}$
                            &$1.64^{+0.31}_{-0.34}$
                          & $1.50^{+0.59}_{-0.34}$\\[1.5ex]
                          
       \texttt{zxipcf} &$ N_{\rm i}^{\rm b}$
			&--
			& $5.1^{+15.2}_{-4.7}$
			&--
                          &$63.8^{+31.1}_{-14.1}$ \\
                          
                          &$\log\xi$
			&--
			& $<3.05$
			&--
                          &$<2.66$ \\
                          
  \texttt{cutoffPL2} & $\Gamma_2$
  			&--
			&  $2.20^{+0.24}_{-0.30}$
			&--
			& $1.72^{+0.44}_{-0.23}$\\
			
			& $E_{\rm cut}$~(keV)
			&--
			&150 (fix)
			&--
			&150 (fix)\\
			
			&$N_{\rm PL}^{\rm c}$
			&--
			&$0.62^{+0.19}_{-0.17}$
			&--
			&$<0.15$\\[1.2ex]

   $\chi^2$/d.o.f. & & 11.52/13& 15.16/9   & 9.26/13 & 17.35/9\\
   \hline\hline

  \end{tabular}
\end{center}
   	{\small
	$^{\rm a}$ The errors refer to 90\% confidence ranges. \\
	$^{\rm b}$ Equivalent hydrogen column density in  $10^{22}$ cm$^{-2}$. \\
         $^{\rm c}$ The \texttt{pexmon} normalization at 1 keV, in units of $10^{-2}$~photons~keV$^{-1}$~cm$^{-2}$~s$^{-1}$~at 1 keV.\\
          $^{\rm d}$ The \texttt{cutoffPL} normalization at 1 keV, in units of $10^{-2}$~photons~keV$^{-1}$~cm$^{-2}$~s$^{-1}$~at 1 keV.\\
          $^{\rm e}$ The \texttt{reflionx} normalization in units of $10^{-6}$.}

\end{table*}
\renewcommand{\arraystretch}{1}

Following the recipe of the C3PO method, 
the data distribution in each CCP was fitted with one straight line, 
expressed by 
\begin{equation}
y = Ax + B,
\end{equation}
in which the slope $A$ and the offset $B$ were both left free. 
Regressions in the fits were performed 
by the Bivariate Correlated Errors and intrinsic Scatter (BCES) algorithm (Akritas \& Bershady 1996) 
to consider both $x$ and $y$ errors. 
As shown in Table 2, the linear fits are all acceptable, and 
the obtained slopes and offsets in the 2005 CCPs are indeed 
larger than those in 2009.  

The value of $B$ in eq.(1) would mean the stationary signal in this band, if $x$, the reference band 
signal, eventually vanish. However, $x$ is in reality considered to have an unknown intensity floor, 
$C$, that represents the non-varying component in the reference band. 
Then, eq. (1) can be rewritten as 
\begin{equation}
y = A(x-C)+B',
\end{equation}
with  
\begin{equation}
B'=B+A C. 
\end{equation}
Since $AC$ is always positive, $B'$ takes a larger positive value than $B$ in any band. 
The quantity  $C$ has an uncertainty over the range of $0 \leq C \leq C_{\rm max}$, 
where $C_{\rm max}$ is the maximum floor allowed by the  data, which is equivalent to 
the minimum count rate recorded in the reference band; $C_{\rm max} \sim 0.03$ cnt s$^{-1}$
in 2005, and $\sim 0.04$ cnt s$^{-1}$ in 2009. An important issue in our subsequent analysis
is how to deal with this uncertainty.

\subsection{A variable spectrum in the 2--45 keV band}

The 2--45 keV variable spectrum can be constructed   
by multiplying the slope $A$ of eq. (2) by $x_0 - C$, 
where $x_0\sim0.04$ cnt s$^{-1}$ in 2005  
and $\sim 0.1$ cnt s$^{-1}$ in 2009 are the average count rate 
in the 1.4--1.7 keV reference band, 
and dividing by the corresponding energy interval. 
The results for $C=0$ and $C=C_{\rm max}$ are plotted 
in Figure 3 and Figure 5 (both in green), respectively, in a form of ratios to a $\Gamma=2$ PL. 
As the value of $C$ becomes larger, the normalization of the variable spectrum decreases 
and becomes minimum at $C=C_{\rm max}$. 
This is simply because the entire spectrum scales with $x_0-C$. 

Because the variable spectrum keeps its shape as $C$ is varied, 
below we analyze the case with $C=0$. 
Because of the obvious resemblance to the difference spectra (Fig. 3), 
we fitted the variable spectra with the same \texttt{model\_v = wabs0 * zxipcf  * cutoffPL0} 
as defined in \S3.2, 
and obtained results  as shown in Figure 3 and Table 1.   
The fits were both successful 
with all parameters consistent with those in the fits 
to the difference spectra, except for the normalization.
Thus, as expected from the difference spectrum analyses, 
the variable components have been reproduced with a single PL with the ionized absorption.

\subsection{A stable component in the 3--45 keV band}

The C3PO method is powerful to determine the non-varying component, as well as the variable part. 
This can be carried out by dividing the values of $B'$ in eq. (2) by the corresponding energy intervals. 
However, unlike the case of \S4.2, both intensities and 
spectral shapes of the derived stable components are sensitive to 
the intensity floor of the reference band, $C$; the higher $C$ becomes, 
the softer and brighter the stable spectrum becomes due to the addition of the $AC$ term
to the value of $B$ in eq. (3). As the two extreme cases, 
the stable components for $C=0$ and $C=C_{\rm max}$ are shown in Figure 5. 
All of them exhibit hard continua, and an intense Fe-K$\alpha$ line at 6.33 keV (6.40 keV 
in the rest frame). This energy, together with an insignificant width 
($\sigma < 190$ eV) when fitted locally with a gaussian, means that the line 
is mostly coming via fluorescence from distant, 
nearly-neutral matter. 
Hereafter, we call the stationary spectra in 2005 with $C=0$ and $C=C_{\rm max}$ 
the 2005 min and the 2005 max cases, respectively. In the same way, the 
stationary spectra in 2009 with $C=0$ and $C=C_{\rm max}$ are called the 2009 min 
and the 2009 max cases, respectively. 

To interpret the stationary spectra, 
reflection from neutral and/or ionized materials is considered the most natural, 
because reflected signals will be less variable than the primaries due to 
large distances to the reflectors or other effects, and will inevitably be accompanied by iron 
fluorescence lines. 
Thus, first we fitted the spectra with an absorbed neutral-disk reflection model,  
\texttt{wabs0 * pexmon}, where \texttt{wabs0}  is the same absorption as introduced in \S3.2, 
with the column density $N_{\rm H0}$ fixed to the values obtained in \S4.2, and 
 \texttt{pexmon} in XSPEC12 represents 
cold reflection, which consists of a Compton-scattered continuum and self-consistent 
Fe and Ni fluorescence lines (Nandra et al. 2007). 
Furthermore, 
the photon index $\Gamma_{\rm ref}$ of the primary continuum in \texttt{pexmon} 
was fixed at 2.12 in 2005, and 1.75 in 2009,  
so as to be consistent with $\Gamma_0$ of the C3PO-derived 
variable spectra (\S4.2). 
The cutoff energy of the primary continuum, the inclination, the Fe abundance 
and the redshift in \texttt{pexmon} were fixed at 150 keV, $60^{\circ}$, 1 Solar, 
and 0.00885, respectively, while the normalization was left free. 
However, as shown in Table 3, the fits were all unsuccessful, with very large $\chi^2$ values. 
These are mainly because the neutral disk reflection is too hard to 
explain the soft energy part of these stationary spectra, 
leaving positive residuals therein. 

\begin{figure*}[t]
\epsscale{1}
\plotone{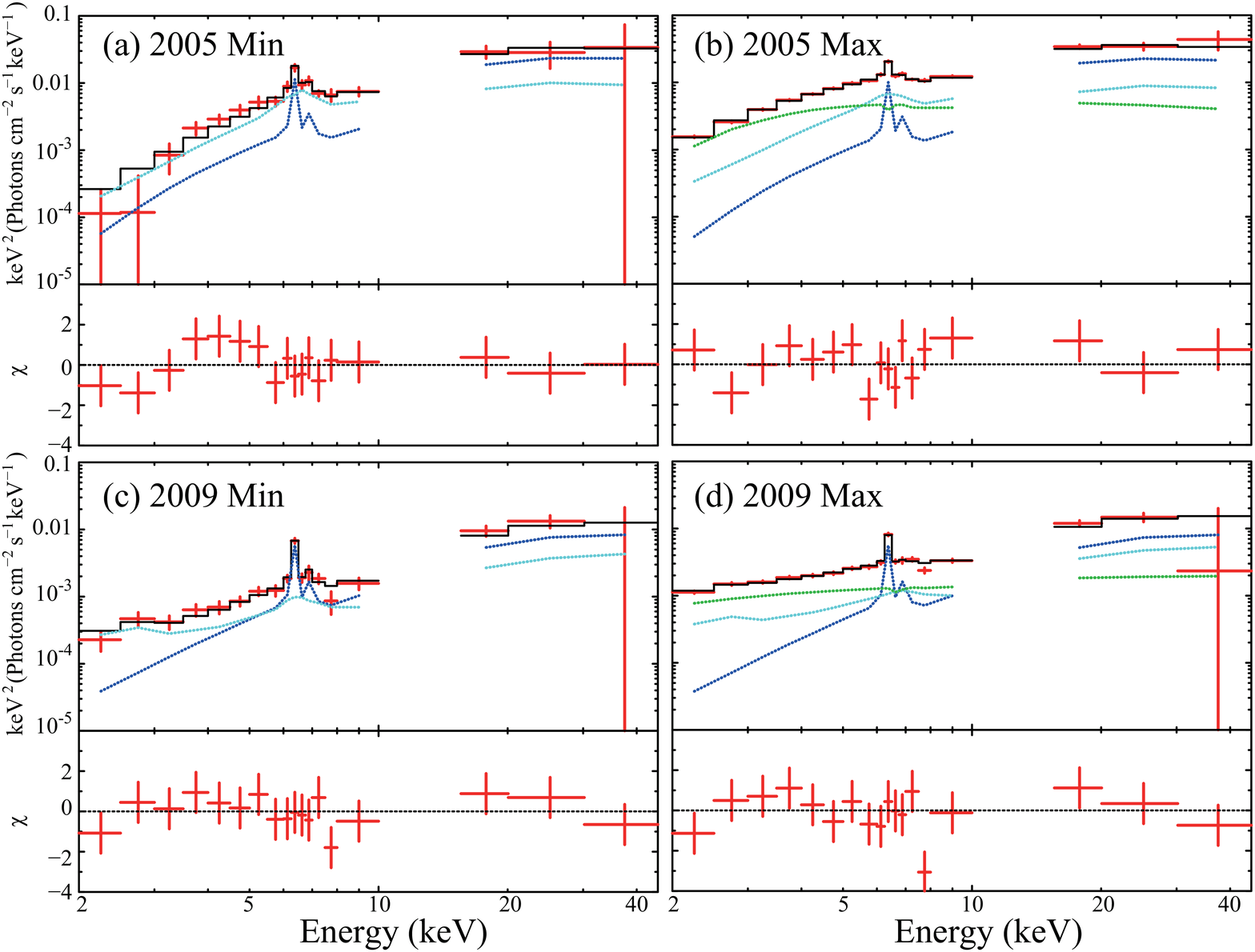}
\caption{Same as Figure 7, but the absorbed cutoff PL model is replaced with a 
relativistically blurred ionized reflection (cyan). That is, 
the utilized model is 
\texttt{wabs * (pexmon + kdblur * reflionx + zxipcf * cutoffPL)} for the 2005 max case, 
while \texttt{wabs * (pexmon + kdblur * reflionx)} for the others. }
\end{figure*}

To improve the fits particularly in the soft energy bands, 
second, we changed the neutral-disk reflection model into an ionized-disk one, 
\texttt{reflionx} (Ross \& Fabian 2005), and fitted the same stable spectra 
with \texttt{wabs0 * reflionx}. 
The \texttt{reflionx} parameters were treated in the same way as those for 
the preceding \texttt{pexmon} case, with an additional free parameter, 
the ionized parameter $\xi$. 
As shown in Table 3,  the fits were somewhat improved, but still unsuccessful 
with $\chi^2$/d.o.f. $\gtrsim 2.4$, except for the 2009 min case. 
This is because the positive residuals still remain in the soft band 
or in the Fe-K$\alpha$ band. 
The soft-band residuals could be reduced by increasing $\xi$, 
but then, the narrow Fe-K$\alpha$ core becomes difficult to reproduce. 
Thus, a single reflection component, either neutral or ionized, 
cannot reproduce the non-variable spectra.

As a third step, we combined  the neutral- and ionized-disk reflection models, 
and fitted the stable components with \texttt{wabs0 * (pexmon + reflionx)}, 
under the same parameter conditions as before. 
As shown in Table 3 and Figure 6, 
the fit remained successful in the 2009 min case, while still unsuccessful in the other cases. 
Although the positive residuals in the soft energy band seen in the previous fits became smaller, 
the concave shape of the stable component (in particular in 2005) in the 
2--6 keV range cannot be explained by reflection models which have 
power-law like or even concave shapes in this band.

To enable the model to have more convex shapes as required by the data, 
we replaced the \texttt{reflionx} component with an empirical absorbed 
cutoff-PL model, denoted by \texttt{cutoffPL1}, and fitted
the stationary spectra with a model of the form 
\texttt{model\_s $\equiv$ wabs0 * (pexmon + wabs1 * cutoffPL1)}. 
The column density $N_{\rm H1}$ of the newly introduced absorption factor \texttt{wabs1} 
was left independent of the $N_{\rm H0}$ parameter in \texttt{wabs0}, 
and the slope $\Gamma_1$ of \texttt{cutoffPL1} was also set 
separate from $\Gamma_{\rm ref}$ of the
primary continuum in \texttt{pexmon}, 
but the  cutoff energy of \texttt{cutoffPL1} was fixed at 150 keV like in \texttt{pexmon}. 
As shown in Table 4 and Figure 7, 
the fits have become successful, except for the 2005 max case which 
still had $\chi^2$/d.o.f $> 8$ (not given in Table 4). 

\renewcommand{\arraystretch}{1}
\begin{table*}[h]
 \caption{Parameters obtained by fitting simultaneously the time-averaged spectrum, 
together with the variable and stable spectra 
 derived by the C3PO method$^{\rm a}$.}
 \label{all_tbl}
 \begin{center}
  \begin{tabular}{ccccc}
   \hline\hline
   Component & Parameter &  2005 & 2009\\
         \hline
     \texttt{wabs0} &$ N_{\rm H0}^{\rm b}$
			& $3.3 \pm 0.4$
			& $0.8^{+0.2}_{-0.4}$\\

     \texttt{zxipcf} &$ N_{\rm i}^{\rm b}$
			&$38.5^{+7.4}_{-6.1}$
                          &$<2.4$ \\
                          
                          &$\log\xi$
			&$3.07 \pm 0.06$
                          &$>4.19$ \\
                         
  \texttt{cutoffPL0} & $\Gamma_0$
  			& $2.21 \pm 0.15$
			& $1.72^{+0.08}_{-0.12}$\\
			
			& $E_{\rm cut}$~(keV)
			&  \multicolumn{2}{c}{150 (fix)}\\
			
			&$N_{\rm PL}^{\rm c}$
			&$1.37^{+0.28}_{-0.32}$
			&$3.18^{+0.49}_{-0.57}$\\
\texttt{gaussian}& $N_{\rm gauss}^{\rm d}$
			&$-6.88^{+1.79}_{-1.90}$
			&$> -2.98$\\[1.5ex]		

     \texttt{gsmooth} & $\sigma$~(keV)
     				&$0.059 \pm 0.008$
				&$<0.038$\\

    \texttt{pexmon} & $\Gamma_\mathrm{ref}$%
    			& \multicolumn{2}{c}{=$\Gamma_{\rm PL}$}\\
                          
                          & $E_{\rm cut}$~(keV)
                      & \multicolumn{2}{c}{150 (fix)}\\
                          
                          &$A_{\rm Fe}$~($Z_{\odot}$)
                          & $1.3 \pm 0.5$
                          &$1.2 \pm 0.5$\\
                          
                         &$f_{\rm ref}$~($\Omega/2\pi$)
                          & $2.1^{+0.7}_{-0.4}$
                          &$2.1 \pm 0.5$\\
                          
                         &$I$~(degree)
                          & \multicolumn{2}{c}{60 (fix)}\\
                          
                          & $N_\mathrm{ref}$%
    		& \multicolumn{2}{c}{=$N_{\rm PL}$}\\

    \texttt{wabs1} &$ N_{\rm H1}^{\rm b}$
			&$7.8^{+1.2}_{-1.9}$
			&$<3.7$\\
   
   \texttt{cutoffPL1}     & $\Gamma_1$%
		& $1.10 \pm 0.06$
		&$1.28^{+0.33}_{-0.22}$\\
                      
                      & $E_{\rm cut}$~(keV)
                     & \multicolumn{2}{c}{150 (fix)}\\

                    & $N_\mathrm{PL}^{\rm c}$%
                     &$1.17^{+0.28}_{-0.24}$
                     &$0.18 \pm 0.01$\\[1.5ex]

   $\chi^{2}$/d.o.f. & & 749.1/740 & 499.0/486   \\
      \hline

     \texttt{wabs0} &$ N_{\rm H0}^{\rm b}$
			& $3.8^{+0.2}_{-0.1}$
			& $1.2^{+0.1}_{-0.2}$\\

     \texttt{zxipcf} &$ N_{\rm i}^{\rm b}$
			&$29.1^{+3.6}_{-2.0}$
                          &$<1.15$ \\
                          
                          &$\log\xi$
			&$3.12 \pm 0.06$
                          &$<3.21$ \\
                         
  \texttt{cutoffPL0} & $\Gamma_0$
  			& $2.16^{+0.02}_{-0.03}$
			& $1.85 \pm 0.04$\\
			
			& $E_{\rm cut}$~(keV)
			&  \multicolumn{2}{c}{150 (fix)}\\
			
			&$N_{\rm PL}^{\rm c}$
			&$1.20^{+0.31}_{-0.26}$
			&$0.36 \pm 0.02$\\
			
\texttt{gaussian}& $N_{\rm gauss}^{\rm d}$
			&$-6.30^{+2.00}_{-1.89}$
			&$> -3.96$\\[1.5ex]	

     \texttt{gsmooth} & $\sigma$~(keV)
     				&$0.053^{+0.009}_{-0.010}$
				&$<0.037$\\

    \texttt{pexmon} & $\Gamma_\mathrm{ref}$%
    			& \multicolumn{2}{c}{=$\Gamma_{\rm PL}$}\\
                          
                          & $E_{\rm cut}$~(keV)
                      & \multicolumn{2}{c}{150 (fix)}\\
                          
                          &$A_{\rm Fe}$~($Z_{\odot}$)
                          &$1.0^{+0.3}_{-0.4}$
                          &$1.1^{+0.3}_{-0.2}$\\
                          
                         &$f_{\rm ref}$~($\Omega/2\pi$)
                          & $2.4 \pm 0.3$
                          &$2.3 \pm 0.1$\\
                          
                         &$I$~(degree)
                          & \multicolumn{2}{c}{60 (fix)}\\
                          
                          & $N_\mathrm{ref}$%
    		& \multicolumn{2}{c}{=$N_{\rm PL}$}\\

    \texttt{kdblur} &$q$
			&$6.2 \pm 0.6$
			&$<4.0$\\
			
			&$R_{\rm in}$
			&$1.7^{+0.3}_{-0.5}$
			&$9.4^{+4.7}_{-6.5}$\\
   
   \texttt{reflionx}     & $\Gamma_{\rm ref}$%
		& \multicolumn{2}{c}{=$\Gamma_{\rm PL}$}\\
                      
                      & $\xi$  (erg cm s$^{-1}$)
                     &$<12.3$
                     &$67.4^{+124.8}_{-33.8}$\\

                    & $N_\mathrm{PL}^{\rm e}$%
                     &$81.34^{+2.13}_{-6.81}$
                     &$1.14^{+0.31}_{-0.25}$\\[1.5ex]

   $\chi^{2}$/d.o.f. & & 860.04/739 & 491.43/485   \\
\hline\hline

  \end{tabular}
\end{center}
   	{\small
	$^{\rm a}$  The errors refer to 90\% confidence ranges.\\
	$^{\rm b}$ Equivalent hydrogen column density in  $10^{22}$ cm$^{-2}$. \\
         $^{\rm c}$ The \texttt{cutoffPL} normalization at 1 keV, in units of $10^{-2}$~photons~keV$^{-1}$~cm$^{-2}$~s$^{-1}$~at 1 keV.\\
         $^{\rm d}$ The Gaussian normalization in units of 
         $10^{-6}$~photons~keV$^{-1}$~cm$^{-2}$~s$^{-1}$.\\ 
       $^{\rm e}$ The \texttt{reflionx} normalization in units of $10^{-6}$.
}

\end{table*}


Finally, let us consider how the 2005 max spectrum can be reproduced. 
When $C$ is increased, 
the variable spectrum decreases in normalization according to eq. (2), 
and the stationary one, $B'$ of eq. (3), increases by $AC$. 
It is hence most natural to assume that the stable spectrum contains a fraction 
of the component that constitutes the varying spectrum. 
Therefore, to the above defined \texttt{model\_s}, 
we added \texttt{zxipcf * cutoffPL2} which was found successful in Figure 3 
to represent the variable signals. 
The column density $N_{\rm i}$ and 
the ionized parameter $\xi$ of \texttt{zxipcf}, 
as well as the photon index $\Gamma_2$ of \texttt{cutoffPL2} were left free, 
while the high energy cutoff of \texttt{cutoffPL2} and the redshift of \texttt{zxipcf} 
were fixed at 150 keV and 0.00885, respectively. 
We then fitted the 2005 max stable spectrum with 
\texttt{wabs0 * (pexmon + wabs1 * cutoffPL1 + zxipcf * cutoffPL2)}. 
As a result, the fit has become  acceptable as shown in Figure 7 and Table 4, 
and the parameter values of \texttt{wabs1 * cutoffPL1} 
were found to be essentially independent of $C$ 
(i.e., the same between the min and max cases within errors). 
In addition, $N_{\rm i}$ and $\Gamma_2$ of \texttt{zxipcf * cutoffPL2} became  consistent 
with those in the fits to the variable components. 
These results just reconfirm our prospect that the stationary spectrum with $C=C_{\rm max}$
should be obtained by adding a fraction of the variable spectrum to the $C=0$ stationary one. 

To realize the convex spectral shape, 
a relativistically-blurred reflection component can be considered as an alternative
to \texttt{wabs1 * cutoffPL1} in the previous fits. 
Hence, we incorporated a relativistic kernel, \texttt{kdblur}, 
and fitted the stable components with 
\texttt{model\_{s}' $\equiv$ wabs0 * (pexmon + kdblur * reflionx)}, 
to find that the fits to the 2005 min and 2009 min cases are successful, 
while the others are not. 
Thus, we fitted the stationary components in the 2005 and 2009 max cases  
with \texttt{wabs0 * (pexmon + kdblur * reflionx + zxipcf * cutoffPL2)} as 
an analogy to the previous attempt. 
As a result, the fits became all successful as shown in Figure 8 and Table 4, 
with $N_{\rm i}$, $\xi$, and $\Gamma_2$ all consistent with those in \S4.2. 
Therefore, not only \texttt{model\_s} employing the absorbed cutoff PL component, 
but also \texttt{model\_{s}'} involving the relativistic ionized reflection, remains 
as a candidate to explain the stationary components detected in the two data sets 
with the C3PO method.

\subsection{Time-Averaged Spectrum Analysis}

\begin{figure*}[t]
\epsscale{1}
\plotone{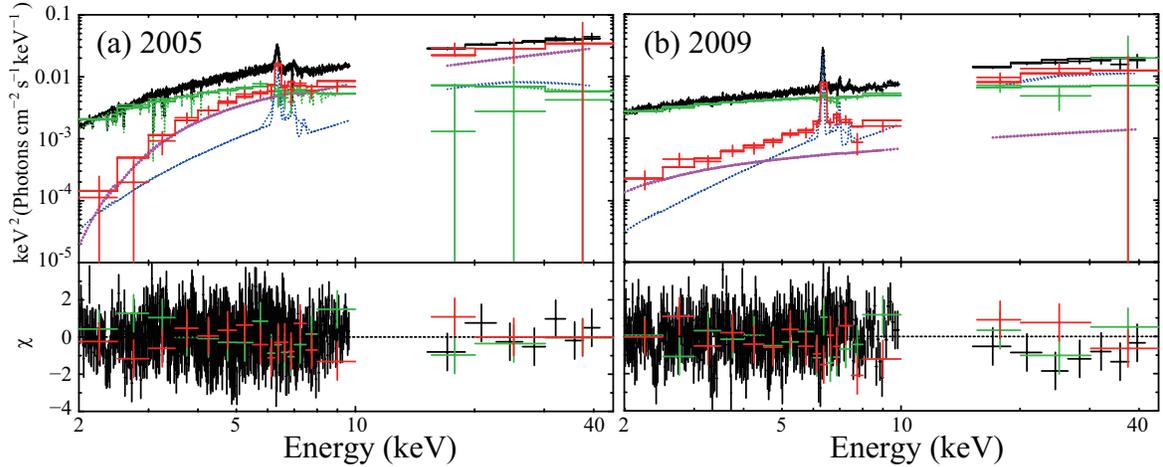}
\caption{Simultaneous fits to the time-averaged spectrum (black), 
the C3PO-derived variable spectrum (green), 
and the stable spectrum (red), shown in $\nu F_{\nu}$ form.  
Panel (a) is for the 2005 data, while (b) is for those of 2009. 
The variable and stable spectra refer to the case of $C=0$. 
The fitted models are \texttt{model\_v + model\_s} (black), \texttt{model\_v} (green), 
and \texttt{model\_s} (blue + purple) for 
the time-averaged, the variable, and the stable spectra, respectively.  }
\end{figure*}

\begin{figure*}[t]
\epsscale{1}
\plotone{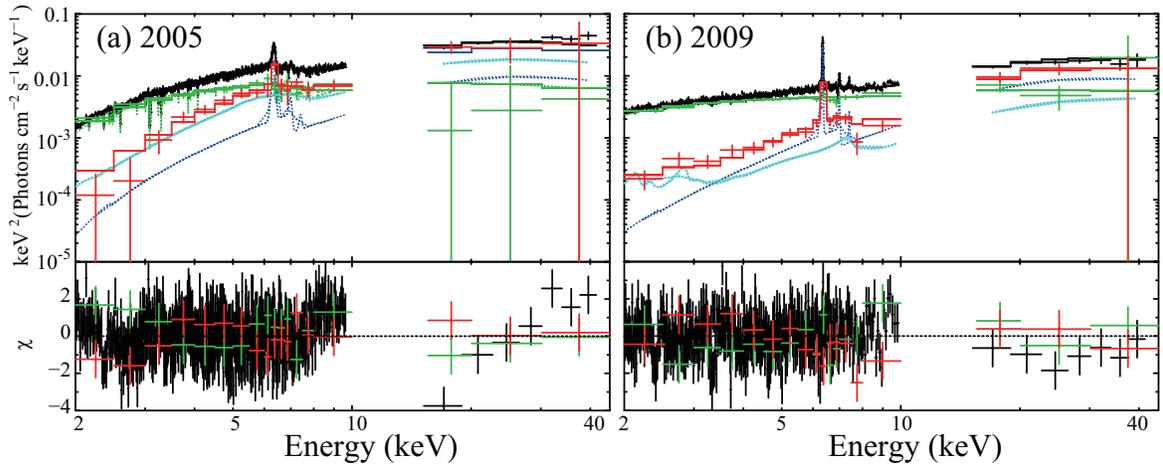}
\caption{Same as Figure 9, but \texttt{model\_s} is replaced with \texttt{model\_s}' (see text).  }
\end{figure*}

In \S4.2, 
the variable spectra were successfully reproduced with 
\texttt{model\_v = wabs0 * zxipcf * cutoffPL0}. 
On the other hand, in \S4.3,  the stationary emission has been explained 
by either \texttt{model\_s} or \texttt{model\_{s}'} (sometimes with the addition of a fraction 
of model\_v). 
We hence expect the time-averaged spectra to be explained with 
\texttt{model\_v + model\_s} or \texttt{model\_v + model\_{s}'}. 
To verify this, we tried simultaneous fits to the time-averaged, the variable, 
and the stationary spectra with the two model forms.  

First, employing the 
\texttt{model\_v + model\_s} combination, 
we  fitted the variable component with \texttt{model\_v}, 
the stable component with 
\texttt{model\_s}, 
and the time-averaged spectrum with \texttt{model\_v + model\_s}. 
Here, a \texttt{gsmooth} model with $\sigma$ left free 
was involved to smear \texttt{pexmon}, 
because  the time-averaged spectra have 
much finer bin size than the stationary components, 
and hence could be subject to some relativistic effects. 
Based on the results obtained in \S4.3, the value of $C$
was chosen to be 0 (i.e., the min case), 
because cases with $C \neq 0$ will be explained simply by changing the intensity of 
\texttt{model\_v}. 
In the fits, the parameter conditions in \texttt{model\_v} and \texttt{model\_s} are the same 
as those in \S4.2 and \S4.3, respectively, 
except the Fe abundance in \texttt{pexmon} left free here. 
As a result, the triplet spectra have been simultaneously reproduced successfully, 
with $\chi^2$/d.o.f.=790.2/741 in 2005, and 500.3/487 in 2009. 
However, at $\sim6.93$ keV, negative residuals still remained, especially in 2005. 
This structure was already reported by Turner et al. (2008) utilizing the \textit{Chandra} HETG. 
We therefore introduced into \texttt{model\_v} a negative gaussian with its center energy fixed at 
$E_{\rm c} = 6.93$ keV and $\sigma$ fixed at 0.01 keV, 
and repeated the fitting using \texttt{model\_v = wabs * (zxipcf * cutoffPL - gauss)} 
plus \texttt{model\_s}. 
As shown in Table 5 and Figure 9, the fit in 2005 was significantly improved to 749.1/740;  
the presence of the additional absorption line at 6.93 keV (6.99 keV in rest frame) 
is significant in 2005. (This feature was not significant in the 2009 spectrum. )  
Thus, the inclusion of the absorbed-PL component has been confirmed 
to give a successful and self-consistent explanation to 
the two \textit{Suzaku} data sets. 

Next,  we replaced \texttt{model\_s} with \texttt{model\_{s}'}  
to examine the relativistic reflection interpretation. 
We again included \texttt{gsmooth} 
with $\sigma$ left free into \texttt{model\_{s}'}, 
and a negative Gaussian with the fixed $E_{\rm c}$ and the fixed 
$\sigma$ into \texttt{model\_v}.  
As shown in Table 5 and Figure 10, 
the simultaneous fitting result in 2009 was successful 
like the case utilizing \texttt{model\_s}, 
while that in 2005 was somewhat worse than the previous fit, giving $\chi^2$/d.o.f.$>1.16$ 
(or $\Delta \chi^2 = 110.9$ against $\Delta \nu = -1$). 
In the latter case, $\chi^2$/d.o.f. values contributed by 
the variable, stationary, and the time-averaged spectrum 
are 14.25/18, 10.15/18, and 835.64/703, respectively. 
Therefore, the chi-square increase is mainly due to the time-averaged spectrum, 
where the model left positive residuals especially in $\sim$8--10 keV (Figure 10a). 
The data thus favor the \texttt{model\_s} interpretation over that with \texttt{model\_s'}. 
However, we cannot completely exclude the latter
only with this fit, since this energy band is easily influenced by the reflection modeling 
(e.g., the choice of inclination). 
Actually, when ignoring the 8--10 keV band, the simultaneous fit to the triplet spectra 
in 2005 with \texttt{model\_s'}
drastically improved to $\chi^2$/d.o.f.$=739.82/699$. 
Hence, the relativistic reflection interpretation 
may still remain a possibility. 
\section{Discussion and Conclusion}

\subsection{Summary of the results}

In addition to the traditional RMS and difference spectrum techniques, 
we employed the C3PO method (Noda et al. 2011b, 2013), 
and successfully decomposed the 2--45 keV emission into the variable and stationary parts. 
Further applying spectral model fits to these two partial spectra, dealing simultaneously with the 
entire time-averaged spectrum, we have revealed that the overall X-ray 
emission of NGC 3516, obtained on the two occasions,  
can be decomposed into the following three components. 

\begin{enumerate}
\item A single cutoff PL (\texttt{cutoffPL0}) with 
$\Gamma_0 \sim2.2$ (2005) or $\sim1.7$ (2009), 
with a relatively low absorption with $N_{\rm H0} \sim 3.3 \times 10^{22}$ cm$^{-2}$ (2005)/
$\sim 0.8 \times 10^{22}$ cm$^{-2}$ (2009). 
In the 2005 data, it is subject to an ionized absorption with $N_{\rm i} \sim 4 \times 10^{23}$ cm$^{-2}$ 
and $\log \xi \sim 3.1$. 
This component, expressed by \texttt{model\_v}, is 
variable on time scales of several hundreds ksec or less, and constitutes 
the entire variable spectrum.
It also explains some fraction of the stationary spectrum, particularly 
if $C$ is chosen to be high.
\item A reflection component from (nearly) neutral materials without relativistic effects 
($\sigma<0.07$ keV in Table 5).  
It includes a narrow Fe-K$\alpha$ emission line, of which the intensity constrained the Fe abundance 
as $\sim 1$ Solar (Table 5). 
It stayed unchanged during the 2005 and 2009 observations, for a gross time 
span of 255 ksec and 544 ksec, respectively. 
Its intensity, if calculated against the 1st component (\texttt{model\_v}), 
means a very large solid angle of reflection, (4--5)$\pi$ (Table 5). This issue is discussion 
in 5.3. 
\item A hard PL (cutoffPL1) with $\Gamma_1 \sim 1.1$, strongly absorbed by  
$N_{\rm H2} \sim 8 \times 10^{22}$ cm$^{-2}$ (in addition to $N_{\rm H0}$), 
which was particularly strong in 2005. 
Although kept stable in the individual observations (like the 2nd component), 
it decreased significantly from 2005 to 2009, so it is variable 
on a much longer time scale 
than the 1st component.  
Together with the second component, 
it constitutes the stationary emission, \texttt{model\_s}. 
This absorbed-PL modeling may be replaced with a relativistically-blurred 
reflection (\texttt{model\_s'})
\end{enumerate}

\subsection{The variable emission}

The variable spectra (the 1st component in \S5.1) 
were well reproduced with a PL-shaped continuum model (\texttt{model\_v}). 
The 2005 and 2009 data show the significantly different PL photon indices, 
$\sim 2.2$ and $\sim 1.7$, respectively. 
The latter is consistent with the typical photon indices of AGNs as long observed 
(e.g., Tucker et al. 1973; Mushotzky 1974), 
and with those of black hole binaries in the low/hard state 
(e.g., Remillard \& McClintock 2006). 
This value is also similar to a theoretical expectation (e.g., Haardt et al. 1993, 1994). 
The 2005 photon index, in contrast, is somewhat larger than the typical value. 
Such steeper PL slopes have often been observed from 
many Narrow Line Seyfert I galaxies 
(e.g, Laor et al. 1994; Boller et al. 1996), and from some  Broad Line Seyfert I galaxies 
including in particular  MCG--6-30-15 (e.g., Miniutti et al. 2007). 
Turner et al. (2011) already reported a similar result on NGC 3516, and 
the present result on the 2005 data reconfirms their report. 

The variable PL spectrum in 2005 is subject to both 
neutral absorption and highly ionized ($\log \xi > 3$) 
absorption, with a column density of $\sim3.3 \times 10^{22}$ cm$^{-2}$ 
and $\sim4 \times 10^{23}$ cm$^{-2}$, respectively. 
Furthermore, an additional absorption line at a rest-frame energy of 6.99 keV was needed. 
This means that the neutral and warm absorbers distribute in 
multiple zones presumably at different distances from the BH
(Turner et al. 2008, 2011). 
These features decreased significantly from 2005 to 2009; 
the distributions of the neutral and warm absorbers must have changed 
during four years, like in other Seyfert galaxies (e.g., Miller et al. 2007).

\subsection{The stationary emission}

As one of the most important results of the present study,
the stationary component, \texttt{model\_s}, 
has been  extracted successfully with the C3PO technique.
Furthermore, it has been further decomposed into 
the distant reflection (the 2nd component in \S~5.1)
and the highly-absorbed hard continuum (the 3rd in \S~5.1).
While the former clearly represents reprocessed signals,
the latter, without sharp spectral features, leaves several possibilities, 
including a relativistically-blurred reflection and an additional primary PL. 
These two components both stayed constant 
for several hundreds ksec during the individual observations,
but varied on longer time scales.
Below, we discuss them separately.

\subsubsection{The distant reflection}

Like in many other AGNs,
this component (the 2nd one in \S~5.1), 
modeled by \texttt{pexmon} in our analysis,
is characterized by the narrow Fe-K emission line
and the hard X-ray hump rising towards the 15--45 keV range.
The relativistic smearing effect working on this component
has been confirmed rather low (Gaussian $\sigma$ in Table 5).
This, together with the line center energy 
(consistent with that of the neutral Fe-K$\alpha$ line)
and the lack of fast variability,
clearly indicates that this component is produced
via reflection/fluorescence in materials located
at large distances ($\gtrsim 5000~R_{\rm g}$ where $R_{\rm g} \sim 10^{12}$ cm
is the gravitational radius) 
from the central BH.
Furthermore, thanks to the C3PO assistance,
the iron abundance of the reprocessor has been 
constrained as $1.3 \pm 0.5$ in 2005, 
and $1.2 \pm 0.5$ in 2009 (in Solar units; Table 5).
That is, the abundance is consistent with 1 Solar.

In the 2009 spectrum,
this component contributed about half the HXD-PIN signals (Fig. 9 b),
regardless of the $C$ value.
As a result, the reflector solid angle $\Omega$,
calculated against the \texttt{model\_v} primary,
becomes very large, $\sim 4\pi$ (Table 5).
However, this apparent discrepancy can be solved
if we consider the fact 
that the source was rather dim in this particular observation,
and that the illuminating primary X-ray flux in its long-term average
must have been considerably higher (as in 2005).
In the 2005 data when the source was brighter,
the value of $\Omega$ is still very large, $\sim 4 \pi$ (Table 5).
This problem is solved in \S5.3.2. 

\subsubsection{The strongly-absorbed hard continuum}

This is a newly identified component (the 3rd in \S~5.1),
and is empirically represented by a strongly absorbed cutoff PL (\texttt{cutoffPL1}).
It is  characterized by a much hard slope ($\Gamma_1 \sim 1.1$)
than the \texttt{model\_v} continuum,
a higher absorption ($N_{\rm H2} \sim 8 \times 10^{22}$ cm$^{2}$ in 2005),
and  the lack of fast variations.
The overall spectral shape change between the two observations 
are mainly attributed to long-term changes of this component.

In previous AGN studies that are based on ``static'' X-ray spectrum analysis, 
the new stationary component  is 
very likely to have been recognized as ``partial absorption'', namely, a fraction of the 
primary emission that reached us though a thick absorber. 
However, our  ``dynamical'' analysis no longer supports this traditional view
(at least in its simplest form), 
since such a partially-absorbed component would 
have the same variation 
characteristics (and the same spectral slopes) as the non-absorbed primary emission, 
and hence would not contribute to the C3PO-derived stationary emission. 
Then, what is the nature of this component?
Generally speaking, it may be explained either as 
a secondary component, or a part of the primary radiation. 

Let us consider the secondary interpretation,
including in particular Compton scattering process 
by some material located at certain distances from the BH.
The lack of time variability may be explained by 
placing the material rather away from the BH,
and the absorbed hard spectral shape may also be reproduced
by adjusting its spatial distribution and ionization state.
However, such a secondary interpretation meets two difficulties.
One is that such signals must inevitably be accompanied by
Fe-K lines (e.g., Sim et al. 2010), and would take up the Fe-K line flux
observed in the C3PO-derived stationary spectra.
This would reduce the Fe-K flux share attributable 
to the distant reflector (the 2nd component in \S 5.1),
and would make its Fe abundance unrealistically low.
The other problem of such a secondary interpretation is 
that the too large a reflection solid angle ($\sim 4\pi$) of the distant reflector
in 2005 (\S 5.3.1) would remain unexplained.
The first problem could be avoided by placing the reprocessor very close to the BH,
as supported by the relatively successful fit with \texttt{model\_s'},
and further invoking the strong ``light bendingh effects" to suppress the variability 
(Miniutti \& Fabian 2004).
However, the second problem will still persist.

The above problem of too large a reflector solid angle 
suggests that the new absorbed PL (the 3rd component in \S 5.1), 
which was particularly strong in 2005,
is in reality a primary emission.
In fact, when this component is considered to 
contribute an additional illuminating X-ray flux,
the solid angle for the distant reflection reduces to 
$\Omega \sim 2.2 \pi$, which is quite reasonable. 
One possible scenario for such an additional primary component
might be provided by the multi-zone Comptonization view developed e.g., 
by Makishima et al. (2008) for Cyg X-1 and Noda et al. (2011b) for Mrk 509 
to explain the soft excess. 
That is, the central engine (Comptonizing coronae) 
of an AGN may well consist of multiple zones 
with significantly different physical parameters. 
Since the photon index of the strongly-absorbed PL 
is flatter  than that of the rapidly varying PL, 
the region emitting the former is considered to have a higher electron temperature 
and/or a higher optical depth than that for the latter. 
However, it is not obvious how this condition can be reconciled with the remaining 
two conditions required for this component, i.e., the slower variation 
and the apparently stronger absorption. 
It may even be conceivable that the low-energy drop of this component 
is in reality not due to absorption, but instead caused, e.g., by 
some non-trivial shapes of the seed photons or by some anisotropy in the hot electron 
distributions in the Comptonizing corona. 
Further examinations of the origin of this component are beyond the scope 
of the present paper, and will be discussed elsewhere.

We thank all members of the Suzaku hardware and software teams and the Science Working Group. 
HN, KM, and SY are supported by the Japan Society for the Promotion of Science (JSPS) 
Research Fellowship for Young Scientists, the Grantin-Aid for Scientific Research (A) (23244024) 
from JSPS, and the Special Postdoctoral Researchers Program in RIKEN, respectively.

\end{document}